\documentclass[useAMS,usenatbib]{mn2e}
\usepackage{graphicx,amssymb,color}
\usepackage[fleqn]{amsmath}
\usepackage[normalem]{ulem}

\title[The fate of exomoons around white dwarfs]
{The fate of exomoons in white dwarf planetary systems}
\author[Payne, Veras, G\"{a}nsicke \& Holman]{
Matthew J. Payne$^{1}$\thanks{E-mail: matthewjohnpayne@gmail.com, mpayne@cfa.harvard.edu}  
Dimitri Veras$^{2}$,
Boris T. G\"{a}nsicke$^{2}$
Matthew J. Holman$^{1}$,
\\
$^{1}$Harvard-Smithsonian Center for Astrophysics, 60 Garden St., MS 51, Cambridge, MA 02138, USA
\\
$^{2}$Department of Physics, University of Warwick, Coventry CV4 7AL, UK
}

\newcommand{\gsim}{\lower.7ex\hbox{$\;\stackrel{\textstyle>}{\sim}\;$}}
\newcommand{\lsim}{\lower.7ex\hbox{$\;\stackrel{\textstyle<}{\sim}\;$}}

\begin{document}
\pagerange{\pageref{firstpage}--\pageref{lastpage}} \pubyear{XXXX} 
\maketitle
\label{firstpage}

\begin{abstract}
Roughly 1000 white dwarfs are known to be polluted with planetary material, and the progenitors of this material are typically assumed to be asteroids. The dynamical architectures which perturb asteroids into white dwarfs are still unknown, but may be crucially dependent on moons liberated from parent planets during post-main-sequence gravitational scattering.
Here, we trace the fate of these exomoons, and show that they more easily achieve deep radial incursions towards the white dwarf than do scattered planets.  Consequently, moons are likely to play a significant role in white dwarf pollution, and in some cases may be the progenitors of the pollution itself.
\end{abstract}

\begin{keywords}
methods: numerical -- 
celestial mechanics -- 
planets and satellites: dynamical evolution and stability -- 
Moon
\end{keywords}

\section{Introduction}\label{SECN:INTRO}

{ 
What is the long-term fate of planetary systems?
Abundant observations reveal substantial clues \citep{farihi2016}, but theoretical explanations are lacking \citep[e.g.][]{veras2016}.
We know that planetary systems exist around evolved stars, including both giant stars\footnote{Sabine Reffert maintains a database at www.lsw.uni-heidelberg.de/users/sreffert/giantplanets.html to which we refer the reader for further references to numerous individual discovery papers.} and white dwarfs \citep[e.g.][]{vanetal2015}.
White dwarf planetary systems feature some combination of atmospheric metal pollution (obtained with spectroscopic absorption lines), orbiting bodies observed through transit photometry, and debris discs with dust and gas from infrared excesses, spectroscopic emission features and Doppler tomography.

One feature common to all of these white dwarf systems is metal pollution (see \citealt*{farihi2016} and \citealt*{veras2016} for review articles). }
White dwarf atmospheres break up accreted material into their constituent chemical elements and then stratify them according to weight \citep{schatzman1945}. 
Consequently, the visible uppermost layers of their atmosphere should contain only a combination of hydrogen, helium and possibly carbon. 
In reality, between one-quarter and one-half of all known white dwarfs \citep{zucetal2003,zucetal2010,koeetal2014} harbour up to 18 elements \citep{kleetal2010,kleetal2011,gaeetal2012,juretal2012,xuetal2013,xuetal2014,wiletal2015}, providing evidence of frequent, on-going accretion of fragmentary planetary material onto white dwarfs.
The details of this chemistry and implications for planet formation are reviewed in \cite{juryou2014}.
In total, about 1000 white dwarfs are known to be polluted with metals \citep{dufetal2007,kleetal2013,genetal2015,kepetal2015,kepetal2016}.

About 40 of these polluted white dwarfs are known to be surrounded by dusty compact debris discs with radial extents of about $1 R_{\odot}$ \citep{farihi2016}. These discs lie within the disruption, or Roche radius, of the white dwarf, and hence are composed of broken up fragments. The distance at which sublimation occurs often lies within this range, producing gas. In eight cases this gas is observable \citep{gaeetal2006,gaeetal2007,gaeetal2008,gaensicke2011,dufetal2012,faretal2012,meletal2012,guoetal2015} and can constrain the disc geometry, which may be eccentric and non-axisymmetric \citep{manetal2016}.

{ A long-term goal has been to combine these detections with one of an orbiting planet \citep{muletal2008,hogetal2009,debetal2011,faeetal2011,steetal2011,fuletal2014,sanetal2016}. A recent success is WD 1145+017 \citep{vanetal2015}, a white dwarf which is both polluted and bears a debris disc, and which }
has recently been shown to also host disintegrating planetesimals. Hourly changes in the shape and depth of the transit light curves of the white dwarf WD 1145+017 have invigorated the post-main-sequence planetary community, motivating a large-scale observational effort \citep{aloetal2016,croetal2016,gaeetal2016,garetal2016,rapetal2016,redetal2016,xuetal2016} and dedicated attempts to explain these observations \citep{guretal2016,veretal2016c}.
The parallel understanding of the state-of-the-art dynamical and theoretical aspects of post-main-sequence planetary science are summarized in \cite{veras2016}.

Traditionally, asteroids have been invoked as the progenitors of the distintegrating planetesimals, debris discs and metal pollution \citep{graetal1990,jura2003,beasok2013}. 
This notion was quantified by \cite{bonetal2011}, \cite{debetal2012} and \cite{frehan2014}, who showed how a planet can perturb an asteroid in the vicinity of a white dwarf, as long as some configurations are avoided \citep{antver2016}. After tidally breaking up \citep{debetal2012,veretal2014a}, the resulting debris is then circularized by stellar radiation \citep[and possibly additional mechanisms, such as gas drag:][]{veretal2015a}, forming a disc which eventually accretes onto the white dwarf \citep{rafikov2011a,rafikov2011b,rafgar2012,metetal2012}.

However, moons liberated during planet-planet scattering in white dwarf systems, a common phenomenon \citep{payetal2016}, may change this general picture in two ways: (1) the moons themselves might accrete directly onto the white dwarf, or (2) the moons can become minor planets and change the efficiency with which asteroids can be perturbed onto the white dwarf.  Regarding the first point, the internal composition of the moons may be similar to those of the asteroid families inferred from the polluted debris. For the second point, a chain of large (moon-sized or planet-sized) bodies may help perturb an asteroid \citep{bonwya2012} into a target as small as a white dwarf, particularly since the inner few au in white dwarf systems will have been cleared out by the increase in size of the star along the giant branch \citep{villiv2009,kunetal2011,musvil2012,adablo2013,norspi2013,viletal2014,staetal2016}.

Here, we track the trajectories of moons which escape from the clutches of their parent planet after the star has become a white dwarf, in order to better understand their role in the pollution process. In Section 2, we describe the planet-based simulations that we use as a foundation for our study. Then, in Section 3, we describe how we add moons into the simulations. Section 4 presents our results, and we conclude in Section 5.

\section{Long Term, Planet-Only Simulations of Planet-Planet Scattering}
\label{SECN:PLANETS}

Simulating the evolution of multiple planets across all phases of stellar evolution is challenging due to computational limitations and the necessity of combining stellar and planetary evolution.  The addition of moons makes this prospect effectively impossible with current technology because they prohibitively decrease the timestep.

Consequently, we must rely on multi-planet simulations \emph{without} moons prior to the white dwarf phase, and then add moons in only at later stages, once it is known that planetary instability is guaranteed on a short timescale.  
Only a few studies have integrated suites of self-consistent multi-planet simulations across the main sequence, giant branch and white dwarf phases.  
\cite{veretal2013a} and \cite{musetal2014} performed two-planet and three-planet simulations respectively.  
However, for computational reasons, both studies modelled stars with main sequence masses of $3M_{\odot}$ or greater. Alternatively, \cite{veras2016b} modelled the fate of Solar system analogues (with a $1.0M_{\odot}$ star), however he had to skip most of the main sequence. The present-day population of white dwarfs corresponds to a progenitor mass range of $1.5M_{\odot}-2.5M_{\odot}$, with a peak at around $2.0M_{\odot}$ (e.g. Fig. 1 of \citealt*{koeetal2014}).  With this mass range in mind, both \cite{vergae2015} and \cite{veretal2016a} simulated systems with four or more planets across all phases of stellar evolution, including the entire main sequence and giant branch phases.

We rely on simulations from both of these studies, where \cite{vergae2015} adopted equal-mass and \cite{veretal2016a} investigated unequal-mass planets within the same system.  
In these simulations, packed systems of planets were integrated for $>10^{10}$ years, with the central star initially being on the main sequence, then passing through the giant branch (an hence losing mass), before settling into the white dwarf phase.
The stellar mass loss causes the planetary semi-major axes to expand, and can trigger late instability.
We use here an ensemble of 119 of these simulations, which all featured planets that remained stable and packed throughout the main sequence and giant branch phase, before suffering their first mutual close encounter along the white dwarf phase.  
These 119 simulations include four- and ten-planet systems, as well as planets with the mass of Jupiter, Saturn, Neptune, Uranus, Earth, and planets with masses down to 0.046 Earth masses.  
Taken together, these 

\citet{payetal2016} investigated a subset of these simulations and determined that the distribution of close approaches between the \emph{planets} of these simulations could efficiently eject moons from a wide range of circumplanetary orbits. In this current investigation, we wish to understand where these moons ultimately go to once they are liberated from circumplanetary orbit. 

\section{Adding Moons to Long Term Simulations of Planet-Planet Scattering}
\label{SECN:MOONS}
%

Here we discuss strategically inserting moons into the systems from \cite{vergae2015} and \cite{veretal2016a} at times and locations which would provide us with the greatest insight.

\subsection{Timescales and timesteps}
\label{SECN:MOONS:BASIC}
The planet-only simulations described in Section \ref{SECN:PLANETS} can be completed relatively rapidly: the timestep required to resolve a system scales with the timescale of the shortest orbit: for planet-only simulations with typical orbits at many AU, orbits typically have periods of a few years, and typical timesteps required to resolve orbits can be many days. 

Adding moons around any planet in such a simulation can cause significant additional computational strain: the orbital period of moons can easily be a few days (or less), requiring timesteps which are measured in hours in order to resolve the system.
This setup ultimately causes typical planet-only simulation run-times to increase by about two orders-of-magnitude. 

A concrete example is the following: numerically integrating a four-planet simulation of the kind illustrated in figure \ref{FIG:DETAIL53} for a short time ($10^4$ years) takes far less than 1 second with only planets in the simulation, but $\sim 200$ seconds when 1 moon is added onto each planet (see Section \ref{SECN:MOONS:METHOD} below for typical moon parameters).
These timescales make it impractical to simulate systems for billions of years, during which moons remain bound to their parent planet.

Two practical considerations allow us to side-step this problem:
(i) We know the approximate time at which planets start to strongly interact: i.e. the time of the first ``close encounter".  Consequently,  we can ignore the system's previous history: the timespan over which planets remain bound and ordered. Our simulation selection guarantees that this span includes the entire main sequence and giant branch lifetimes, and at least some amount of time on the white dwarf phase.  We know that the moons will remain bound to their parent planets during these earlier phases with few exceptions \citep{payetal2013}.

(ii) While the moons are bound, timesteps must remain short, but if the moons become unbound from their parent planet and move onto planet-like orbits, then the simulation timestep can increase by orders of magnitude, allowing simulations to rapidly progress.

\subsection{Insertion point}
\label{SECN:MOONS:METHOD}

In order to identify the first close encounter time, we use a simple, approximate definition of orbit crossing: when, for an adjacent pair of planets, the pericentre of the outer orbit overlaps with the apocentre of the inner orbit.  This method ignores subtleties associated with resonant orbits, but suffices for our purpose. 

After identifying the first close encounter time, we extract the state of the system (masses, positions and velocities of all bodies) at a time $10^6$ years prior to the onset of orbit-crossing. 
We adopt this previous timestep as our new ``time-zero'', and add moons to the simulation at this time in the manner detailed in section\ref{SECN:PROP}.

{
We then integrate the simulations forward, through the first close planet-planet encounter and onwards through the next $10^8$ years of strong planet-planet interactions as the planets, now on crossing-orbits, repeatedly scatter in the manner illustrated in (e.g.) figure 1 of \citet{vergae2015} and figure~\ref{FIG:DETAIL53} of this paper. Further details and dicussion are are provided in sections~\ref{SECN:NEW} and \ref{SECN:DISCUSSION}.

We note that the first close planet-planet scattering encounter (and hence possible moon liberation) should be expected to occur sometime \emph{after} the onset of orbit-crossing.
}

\subsection{Moon properties}
\label{SECN:PROP}
We add one moon to each of the planets in the simulation.  
The moon is integrated as a test-particle with mass $m_{\rm m} = 0$.
The initial semi-major axis of the moon with respect to the planet, $a_{\rm m}$, in units of the parent planet's instantaneous Hill Radius, $r_{\rm H}$, is chosen randomly such that the distribution of the semi-major axes is uniform in log-space in the range $0.04 < a_{\rm m}/r_{\rm H} < 0.4$.  
Although we know from \cite{payetal2016} that moons can be liberated even from orbits with $a_{\rm m}/r_{\rm H} < 10^{-2}$, here we are particularly interested in the \emph{fate} of moons once liberated, rather than the fine details of which moons in particular will be liberated. As such, we choose to focus our attention on moons which occupy orbits with semi-major axes in the range $0.04 < a_{\rm m}/r_{\rm H} < 0.4$: 
Such a range of semi-major axes was chosen to ensure that 
(a) the outer edge of the distribution is just interior to the stability boundary at $\sim 0.5 r_{\rm H}$, and 
(b) the inner edge of the distribution is sufficiently distant from the planet to make the orbital period manageably long.

The inclinations $i_{\rm m}$ (of a moon with respect to its parent planet) were drawn randomly from a uniform distribution and have values all within $1^{\circ}$ of the plane of the planetary orbits (which themselves \emph{initially} had mutual inclinations within about a degree at the start of the simulations, but by the time we insert the moons, have started to excite larger inclinations).  
The small but non-zero inclination guarantees that the systems are fully three-dimensional in their interactions, while the low inclinations prevent any unwanted loss of moons via the Kozai mechanism.
Once liberated from their parent planets, the inclination of the moons becomes highly non-complanar, erasing the memory of their initial plane (see figure \ref{FIG:INC}).

The longitude of ascending node, argument of pericentre and mean anomaly of each moon were all drawn from a uniform distribution between $0^{\circ}$ and $360^{\circ}$.

\subsection{New integrations}
\label{SECN:NEW}
\begin{figure}
\begin{minipage}[b]{\columnwidth}
\centering
\begin{tabular}{c}
\includegraphics[trim = 0mm 0mm 0mm 0mm, clip, angle=0, width=0.9\textwidth]{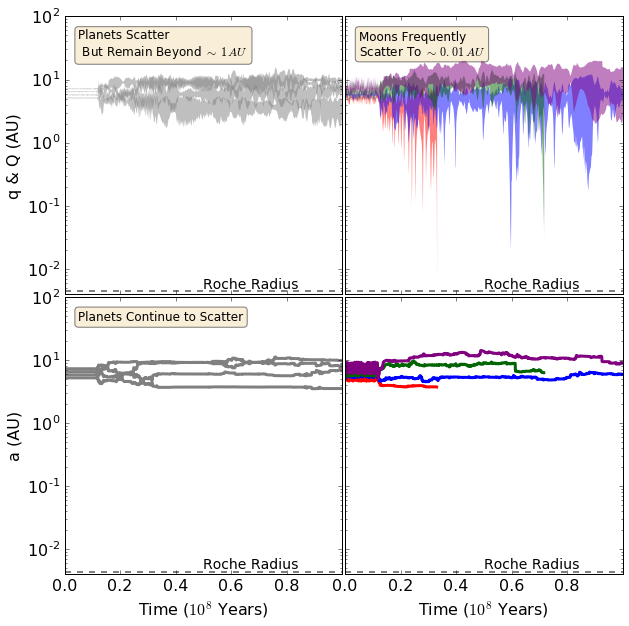}
\end{tabular}
\caption{Example of moon and planet evolution after moons have been liberated.
We plot the semi-major axes (bottom) and pericenters and apocenters (top) of four Earth-mass planets (grey, left) and four (test particle) moons (colours, right).
The semi-major axes are plotted using solid lines, while the pericenters and apocenters are plotted using filled ranges.
The moons are initially bound to the planets, one moon per planet. 
After $\sim\,10^7$ years, planet-planet scattering commences, unbinding all four moons (in this example) from their parent planets and liberating them into white dwarf-centric orbits.
The planets remain relatively distant from the white dwarf (beyond $\sim\,1\,$AU), but the moons are highly scattered, frequently coming within $\sim\,0.1\,$AU of the WD. The red moon comes within a factor of a few of the disruption, or Roche, distance (plotted as a grey dashed line, assuming a density of $3\,$g\,cm$^{-3}$) at closest-approach.
}
\label{FIG:DETAIL53}
\end{minipage}
\end{figure}

The integrations were performed using the Bulirsch-Stoer algorithm from the {\sc Mercury} $N$-Body package of \citet{Chambers99} with an accuracy of $10^{-13}$ and a run-time of $10^8$\,yr. 
Such a run-time is sufficient to guarantee multiple close planet-planet encounters, while remaining computationally tractable for the short timesteps required for bound moons.
We refer the reader to figure~\ref{FIG:DETAIL53} for an example of the multiple close-encounters which can occur during the $10^8$ year simulation, clearly visible as the semi-major axes of the planets repeatedly perform discontinuous jumps in semi-major axis.

Because the integrations took place on the white dwarf phase, it was not necessary to take stellar evolution into consideration. 
Note that unlike main sequence and giant branch stars, white dwarfs do not have winds. 
Hence, orbiting bodies are not affected by stellar mass loss (see Section 4 of \citealt*{veras2016}). 
Objects under 1000km in size would be affected by radiation from the parent star on the giant branch phase \citep{veretal2015b}, but not around a white dwarf unless the object was a boulder (approximately $0.1$m) or smaller \citep{veretal2015a} or was outgassing significant volatiles \citep{veretal2015c}. 
Here, we consider just point-mass gravitational dynamics.

{ 
We note that in Payne et al. (2016) we illustrated the range of moon semi-major axes (with respect to their parent planets) from which planet-planet scattering can efficiently cause moons to be liberated from their parent planet into heliocentric orbits. 
A future study will provide additional details of the exact dependencies of this liberation on the properties of the scattering planets as well as the initial orbits of the moons. 
However, the point of the integrations in this current study is to demonstrate what happens to the moons once they are liberated from their parent planets. 
We emphasize that once the moons are liberated into heliocentric orbits, their heliocentric orbits, by definition, must be planet-crossing, hence their subsequent evolution will be chaotic, driven by multiple hard-scattering events (during close-approaches with the massive planets in the system) which naturally wipe all memory of the moons' initial circumplanetary orbits. 
We wish to use these simulations to understand the range of distances through which this hard-scattering drives the moons, and to understand whether this is different to that of the planets from which they are liberated.
}

%

%
\begin{figure}
\begin{minipage}[b]{\columnwidth}
\centering
\begin{tabular}{c}
\includegraphics[trim = 0mm 0mm 0mm 0mm, clip, angle=0, width=0.9\textwidth]{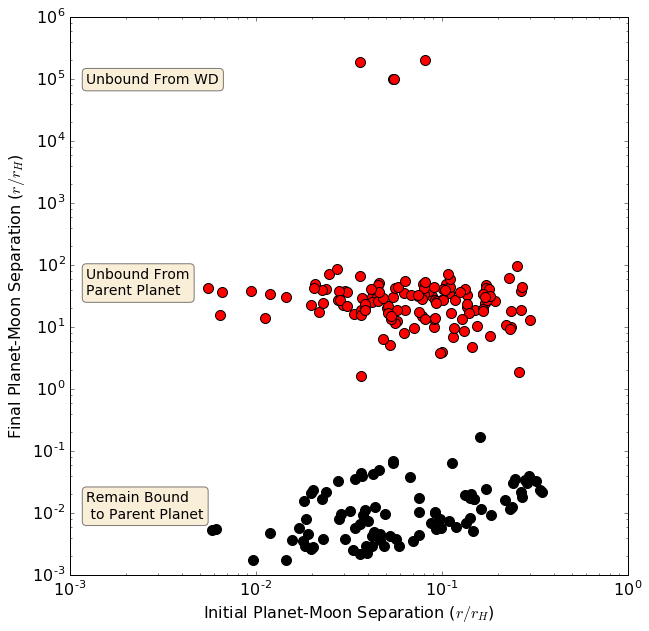}
\end{tabular}
\caption{Distribution of separations between moons and their parent planets.
On the horizontal axis we plot the initial separation and on the vertical axis we plot the final separation after $10^8\,$yr of simulation.
We see that the moons split into two main groups, with the moons at the bottom (black dots) being those that remain bound to their parent planet, while those at the top (in red) have become unbound from their parent planet and have moved into heliocentric (or white dwarf-centric) orbits (of which three have been completely ejected from the white dwarf system). 
It is highly plausible that with increased simulation time, additional moons will become unbound, as (a) many moons remain bound at a wide range of $r/r_{\rm H}$ values, and (b) planet-planet encounters continue over Gyr timescales (see figure\ref{FIG:DETAIL53} and figure 1 of \citet{vergae2015}).
}
\label{FIG:UNBOUND}
\end{minipage}
\end{figure}

\section{Numerical integration results}
\label{SECN:MOONS:RESULTS}
First consider the evolution of a single system, as in Fig. \ref{FIG:DETAIL53}.  In that system, four Earth-mass planets orbiting a $1.5M_{\odot}$ main sequence star remain stable until the white dwarf phase. 
About 1 Myr before the planets first cross orbits we added one moon to each planet and then integrated the system forward for $10^8$ yr.  
What is plotted is the subsequent evolution. 
The moons were initially placed at distances of $a/r_{\rm H} = 0.13,0.24,0.06$ and $0.05$ from the planets (one moon per planet) in planet order from the star.

After about 10 Myr, the close encounters (between planets) result in strong scattering events which (in this example) liberate all four moons within a short period of time.  
These moons become new minor planets themselves, orbiting the white dwarf instead of any of the extant planets.  
Now the system effectively has eight planets.  
Three of the former moons (blue, red and green) change their orbits in such a way as to achieve pericentres below 0.1 AU, an order of magnitude less than the pericentres of the Earth-mass planets.

Now consider the results from our ensemble of 119 systems.  
Not all moons are liberated from their parent planets and stay in the system.  
figure \ref{FIG:UNBOUND} reveals the different possible qualitative outcomes, as a function of separation from parent planet.  
{ Because many moons remain bound, over a wide range of $r/r_{\rm H}$, at $10^8$ years, they will be susceptible to liberation if further planet-planet scattering encounters occur \citep{payetal2016}. 
As planet-planet scattering around WDs can occur over Gyr timescales (far longer than examined here: see figure 1 of \citet{vergae2015} for examples), many more close encounters will occur, making it extremely likely that yet more moons will be liberated from their parent planets at future times.
We note that the fine details of an individual planet-planet scattering encounter (e.g. the distance of close approach) are essentially stochastic, so over time, a greater range of planet-planet encounter parameters will be explored, leading to an increased probability of more tightly bound moons being ejected as time goes on. 
However, once the moons are ejected from their parent planet, all memory of their initial conditions is erased by strong scattering between the liberated moons (now minor-planets) and the large planets whose orbits they cross. 
}

\begin{figure}
\begin{minipage}[b]{\columnwidth}
\centering
\begin{tabular}{c}
\includegraphics[trim = 0mm 0mm 0mm 0mm, clip, angle=0, width=0.9\textwidth]{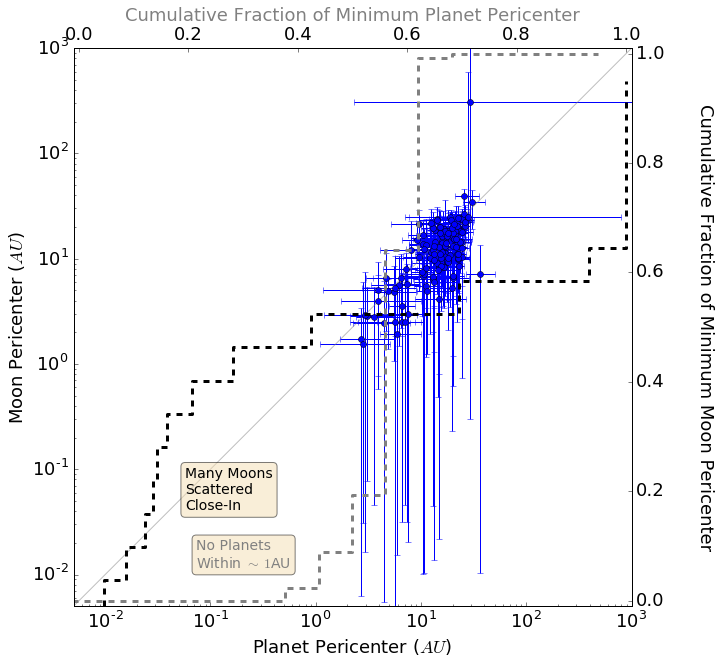}\\
\includegraphics[trim = 0mm 0mm 0mm 0mm, clip, angle=0, width=0.9\textwidth]{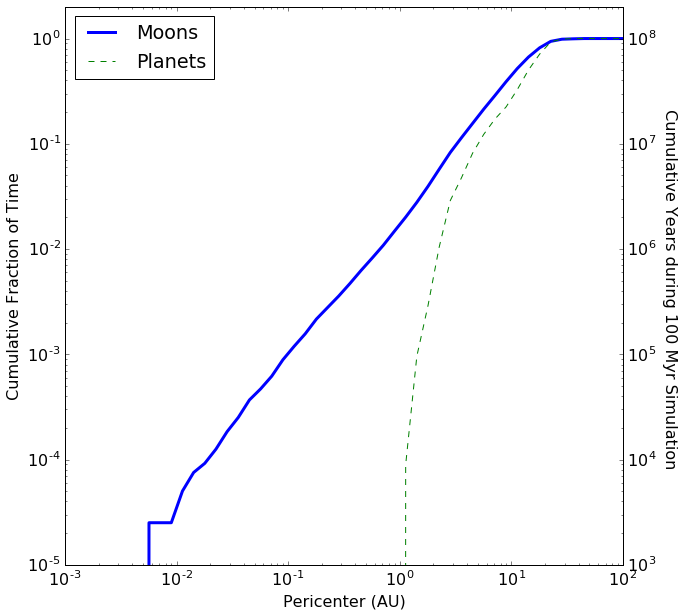}
\end{tabular}
\caption{{\bf Top: }Distribution of pericenters for the moons from figure \ref{FIG:UNBOUND} which are unbound from their parent planets but still bound to the white dwarf system.
On the horizontal axis we plot the pericenters of the \emph{parent planets}, while on the vertical axis we plot the pericenters of the \emph{moons}.
The points indicate the \emph{median} pericenter values for each, while the error bars indicate the minimum and maximum pericenter ever achieved by the object. 
We see that for these simulations the planets never come inside $\sim\,1\,$AU, while the moons frequently come in to $\sim\,0.01\,$AU (a few Roche radii).
The dashed lines provide cumulative-histograms of the \emph{minimum} pericenter distributions for the planets (gray) and moons (black), with the scales being on the right and top axes respectively.
{\bf Bottom: } Cumulative fraction of time with a given pericenter. The moon (thick blue line) spend a much greater fraction of time at small pericenters than do the planets (thin dashed line). 
}
\label{FIG:PERI}
\end{minipage}
\end{figure}

The moons of greatest interest are those which have escaped from their parent planet, but not from the white dwarf system.  
The minimum orbital pericentres of these moons are perhaps the most consequential parameters for white dwarf pollution.  
Hence, in Fig. \ref{FIG:PERI}, we illustrate the distribution of pericenters, showing that $\sim15\%$ of moons come within $0.1$ AU, and $\sim5\%$ come within $10^{-2}$ AU. 
Also plotted are the minimum orbital pericentres of the parent planets: comparing both pericentres illustrates that moons are much more effective at achieving intrusive radial incursions towards the white dwarf than moons, this is the key result of this work.
At the bottom of Fig. \ref{FIG:PERI} we plot the cumulative histogram of the time spent with a given pericenter over all simulations. 
The planets (thin dashed line) spend no significant amount of time inside $a\,AU$, while the moons effectively display a power-law dependence, spending $\sim10^5$years inside $0.1\,$au and $\sim10^4$years inside $0.01\,$au out of this $10^8$ year simulation.

\begin{figure}
\begin{minipage}[b]{\columnwidth}
\centering
\begin{tabular}{c}
\includegraphics[trim = 0mm 0mm 0mm 0mm, clip, angle=0, width=0.9\textwidth]{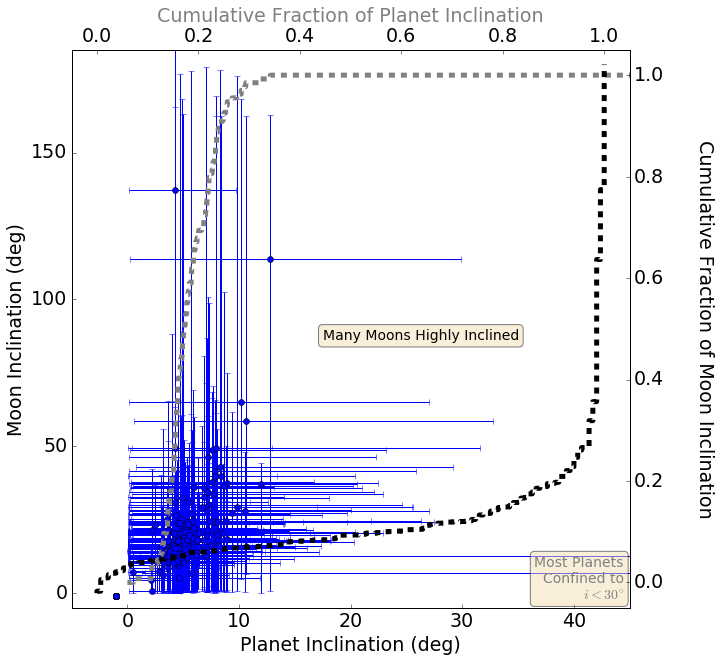}
\end{tabular}
\caption{Distribution of inclinations for the moons from figure \ref{FIG:UNBOUND} which are unbound from their parent planets but still bound to the white dwarf system.
On the horizontal axis we plot the inclination of the \emph{parent planets}, while on the vertical axis we plot the inclination of the \emph{moons}.
The points indicate the \emph{median} inclination values for each, while the error bars indicate the minimum and maximum inclination ever observed for the object in the simulations. 
We see that for these simulations the planets generally remain confined within $\lsim\,30^{\circ}$, while the moons frequently become highly inclined or even retrograde.
The dashed lines provide cumulative-histograms of the inclination distributions for the planets (gray) and moons (black), with the scales being on the right and top axes respectively.
}
\label{FIG:INC}
\end{minipage}
\end{figure}

Even slight initial inclinations of less than a degree can, after scattering, generate inclinations spanning the entire range \citep[e.g.][]{verarm2005,rayetal2010,matetal2013,lietal2014}. 
Consequently, in Fig. \ref{FIG:INC}, we compare the inclinations of the scattered planets and scattered moons.  
The figure demonstrates that moons easily achieve higher inclinations, including near-polar and retrograde inclinations. 

{ 
As seen in figure~\ref{FIG:DETAIL53}, as the white dwarf continues to cool, the system will continue to dynamically evolve, as the planets and liberated moons occupy crossing-orbits.
The chaotic evolution driven by hard scattering between planets and liberated moons causes the power-law distribution of pericenters in the bottom of figure \ref{FIG:UNBOUND}. 
This means that over longer time periods, it becomes increasingly likely that some moons will eventually come close-to, and possibly collide-with, the white dwarf.
}

%
%

\section{Moons and Pollution}\label{SECN:POLLUTION}
 {
In section~\ref{SECN:MOONS:RESULTS} we demonstrated that liberated moons which scatter around the system in the manner depicted in figure~\ref{FIG:DETAIL53} can spend a non-trivial amount of time within $0.1$ au or even as close as $0.01$ au.
}

\subsection{Tidal Disruption}
\label{TIDES}
 {
At close pericenter approaches, moons may be subject to both tidal interactions with the star, and radiative effects, neither of which were modelled in section~\ref{SECN:MOONS:RESULTS}. 

The effect of tides is strongly dependent on the internal composition of the exo-moons. 
Differing rheologies can cause the circularization timescale to vary by orders of magnitude \citep{HH2014} and therefore must be treated on a case-by-case basis. 
Simple tidal models such as the constant geometric lag model have proven false \citep{EM2013} and cannot be used for quantification.

However, it is possible that radiation and tides might act to circularize and shrink the orbit, if not destroy the moon through overspinning \citep{veretal2014b}.
To demonstrate the plausibility of this scenario, we consider the tidal disruption radius,$r_{\rm c}$, described in equation (2) of \citet{veretal2014a}:
\begin{equation}
\frac{r_{\rm c}}{R_{\odot}} = C \left(\frac{M_{\rm WD}}{0.6M_{\odot}}\right)\left(\frac{\rho}{3\,g\,cm^{-3}}\right)^{-1/3}
\end{equation}
where C is a constant ranging from about 0.85 to 1.89
\citep{beasok2013}, $M_{\rm WD}$ is the mass of the WD, and $\rho$ is the assumed density of the moon.
For $C=1.89$ and $\rho=1.8\,g\,cm^{-3}$ (the same as Callisto), then 
\begin{equation}
r_{\rm c}\,\approx\,2.2R_{\odot}\,\approx\,0.011 AU.
\end{equation}
As we illustrated in Figure~\ref{FIG:PERI}, moons do indeed get scattered inside this critical radius for tidal disruption. Hence, for the example coefficients chosen, these moons would be tidally disrupted as they passed within $r_{\rm c}$ close to pericenter passage. 

The disruption of a single moon, which must initially be on a highly eccentric orbit in order to enter the disruption sphere, results in a highly eccentric ring of debris \citep{veretal2014a}. 
The subsequent evolution of the particles in this eccentric debris ring is strongly dependent on particle size: \citet{veretal2015a} demonstrate that circularization of these orbits occurs efficiently for fragments in the size range $10^{-5} - 10^{-1}\,{\rm m}$, on timescales many orders of magnitude shorter than the cooling age of the white dwarf, due solely to the radiation effects from the white dwarf acting on the fragments. 
}

\subsection{Scattering of Small Bodies}
\label{SCAT}
\begin{figure}
    \begin{minipage}[b]{\columnwidth}
    \centering
    \begin{tabular}{c}
    \includegraphics[trim = 0mm 0mm 0mm 0mm, clip, angle=0, width=0.9\textwidth]{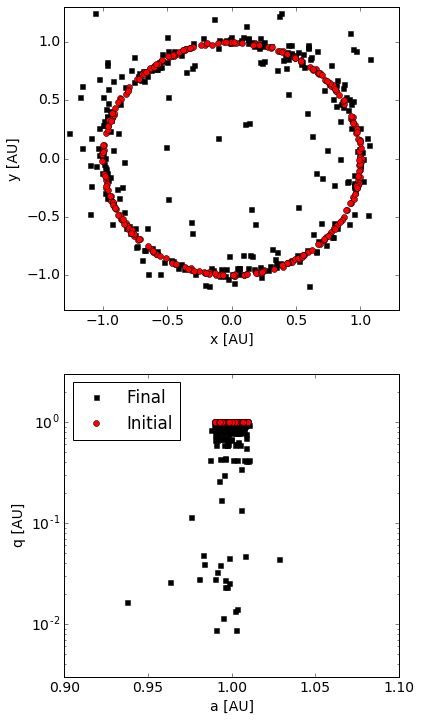}
    \end{tabular}
    \caption{Scattering of interior belt by eccentric moons.
    We add a belt of test particles to the simulation illustrated in Figure~\ref{FIG:DETAIL53}.
    The belt is initially located at $a_{\rm belt}=1.0$ au (red circles).
    The final state of the belt is plotted using black squares. 
    In the top panel we show a Cartesian snapshot of the planet, illustrating the initial conditions and the final positions at $t=10^7$ yrs.
    In the bottom panel we show the $a$ and $q$ values.
    The belt is excited to pericenters $\sim0.01$  au within $10$ Myr. 
    }
    \label{FIG:SCAT}
    \end{minipage}
    \end{figure}

 {
For those moons which do \emph{not} directly encounter the tidal disruption radius of the white dwarf, they may still be able to contribute to the pollution \emph{indirectly} by acting as a perturbing mechanism for smaller bodies to be scattered close to the white dwarf.

If we consider a population of small bodies (analogous to the asteroid belt) distributed in an annulus with semi-major axis $a_{belt}$, then they will initially be unperturbed by the more distant planets, but \emph{may} be perturbed by the either the planets or moons once planetary orbit-crossing commences, exciting the planets and liberating the moons. 
To test this scenario, we inject an annular distribution of test particles with $a_{belt}=1.0\,au$ into the example simulation illustrated in Figure~\ref{FIG:DETAIL53}.
We then integrate the simulation to understand whether the scattered moons disrupt the annulus. 
We integrate the simulations from $t=10^7$ years (just prior to the onset of scattering in figure~\ref{FIG:DETAIL53}) for a further $10^7$ years, and plot the subsequent evolution of the particles in Figure~\ref{FIG:SCAT}.
    }
    
    { 
    In Figure \ref{FIG:SCAT}, the belt is initially located at $a_{\rm belt}=1.0\,au$ (red circles).
    The final state of the belt is plotted using black squares. 
    In the top panel we show a Cartesian snapshot of the belt, illustrating the initial conditions and the final positions at $t=10^7$ yrs.
    In the bottom panel we show the $a$ and $q$ values.
    The belt is excited to pericenters $\sim0.01$ au within $10$ Myr by perturbations from the scattered moons. 
    
    We also repeated the experiment for a belt with $a_{\rm belt} = 0.3\,au$ (\emph{not} illustrated) and found that such a belt was \emph{not} significantly excited by the scattered moons. 
    
    We find that, for certain configurations of small body populations, it is possible for liberated moons to excite the bodies onto orbits which cross the white dwarf's Roche radius.
    
    }

\section{Discussion}\label{SECN:DISCUSSION}
Our results show that liberated moons more easily create a constantly changing dynamic environment than do planets alone.  
Exomoons are regularly perturbed within $10^{-1}$ au or even $10^{-2}$ au of the white dwarf.
{ 
As discussed in section~\ref{SECN:MOONS:RESULTS}, many moons remain bound at $10^8$ years (figure~\ref{FIG:UNBOUND}), and planet-planet collisions continue on Gyr timescales, hence any moons which may happen to remain bound to their parent planet after 100 Myr (the period of time we simulate) may still be ejected at later times.
Moreover, the chaotic nature of the scattering of small bodies (liberated moons on heliocentric orbits) by planets on crossing orbits, naturally means that over time, a greater range of parameter space will be explored, leading to a greater chance of small pericenter orbits being explored by liberated moons. 
}

{

When moons are scattered to very small pericenters (section~\ref{SECN:MOONS:RESULTS}), they may become tidally disrupted (section~\ref{TIDES}).
This is of particular relevance given the non-trivial fraction of simulation time in which we observe moons close to the WD (see figure \ref{FIG:PERI}), suggesting that such interactions could have a significant integrated effect.

Consequently, the moons might represent pollutants themselves.  
E.g. one (speculative) origin for the observed debris in the WD 1145+017 system \citep{vanetal2015} is that a moon is a direct cause of one or more transits.  
If so, then it would likely have been circularized, perhaps through some combination of tides and or gas drag.

Such a scenario escapes one of the significant challenges of explaining observed accretion rates with asteroids, which is that, on average, exo-asteroid belts would need to be about $10^3$ as massive as the Solar system asteroid belt \citep{debetal2012}.  
Moons provide a larger reservoir, as the total mass in moons in the Solar system is about two orders of magnitude larger than the mass in the asteroid belt.  
}

{
Alternatively, a moon could act as a perturbing mechanism for smaller bodies to be scattered close to white dwarfs such as WD 1145+017. 
One mechanism by which this may occur is for a more distant population such as the Kuiper belt.
While it has been demonstrated that such a population can explain accretion rates \citep{bonetal2011}, perturbed Kuiper belt objects have not yet been shown to reach the small target of the white dwarf itself. 
In this respect, liberated moons meandering within 1\,AU may provide a crucial component of the conveyor belt provided by the remaining planets \citep{bonwya2012}.  
Such moons may also increase the efficiency rate of cometary impacts \citep{alcetal1986,verwya2012,veretal2014c,veretal2014d,stoetal2015}, although compositionally comets remain disfavoured \citep{kleetal2010,kleetal2011,gaeetal2012,juretal2012,xuetal2013,xuetal2014,wiletal2015} even for progenitors which are thought to be water-rich \citep{faretal2013,radetal2015}.

An alternative mechanism could be for a small-body population at relatively small semi-major axis (e.g. $0.01-1.0$ au) to be directly perturbed by the moon during its small pericenter incursions.
In section~\ref{SCAT}, we demonstrated that such scattering of interior asteroid/debris belts can indeed occur.
However, the existence of such a population after the significant stellar expansion of the giant branch is highly speculative. 
}

The moons themselves are highly unlikely to significantly affect the orbits of one another because of their small masses, unless they have already reached the white dwarf disruption radius (and have started to break up).  For example, for multiple Ceres-mass and lower co-orbital bodies with orbital periods of just about 4.5 hours, the resulting orbital period deviations due to their mutual perturbations is on the order of seconds \citep{guretal2016,veretal2016b}.  This deviation is observationally relevant for the WD 1145+017 system \citep{gaeetal2016,garetal2016,rapetal2016}, but only because these orbits are so compact.  Mutual perturbations amongst moons are primarily important during the formation of the moons themselves and their subsequent evolution around their parent planet; in our Solar system, there exist several relevant examples (e.g. Io, Europa and Ganymede, and Mimas and Tethys). 
{
Also irrelevant are long-range interactions between planets, as they have been demonstrated \citep{payetal2013} to have little effect on the stability of moons: it is the hard scattering between planets that causes the liberation of moons \citep{payetal2016}.
}
%

\section{Conclusions}\label{SECN:CONC}

With the knowledge that liberating moons from their parent planet during the white dwarf phase is a common process \citep{payetal2016}, here we have tracked the fate of these moons.  We find

\begin{itemize}
    \item Liberated moons can easily become minor planets.
    \item Liberated moons more easily meander about the inner reaches (within 1 au) of a white dwarf system than their parent planets.
    \item The minimum pericentre achieved by the liberated moons, even after just $10^8$ yr, is typically under $10^{-2}$ au.
\end{itemize}

Consequently, the liberated moons may act as \emph{either} a direct source of pollutant material, or the innermost component of a conveyor belt which allows smaller bodies in the system (such as asteroids) to be perturbed onto the white dwarf \citep{bonwya2012}. 
As seen in figure~\ref{FIG:UNBOUND}, many moons remain bound after the $10^8$ yr modelled here, yet planet-planet scattering around WDs continues over Gyr timescales \citet{vergae2015}, providning further opportunity for liberation of moons. 
Hence the prevalence of moons in inner regions is likely to increase as these systems are tracked over longer timespans.

We do not intend this paper to represent an accurate gauge of the fraction of moons that become unbound, but rather as a qualitative assessment of what can happen to moons once they become unbound in white dwarf systems, demonstrating that they can and do go on to repeatedly experience very close-pericentre encounters with the white dwarf.  
Further work will be required to understand the relative importance of liberated moon material compared to other potential sources (asteroidal, cometary, planetary) in polluted white dwarf systems. 


\section*{Acknowledgements}

{ The authors thank the anonymous referee for their comments and suggestions.}
All authors gratefully acknowledge the Royal Society, whose funding (grant number IE140641) 
supported the research leading to these results.  
MJP also acknowledges 
NASA Origins of Solar Systems Program grant NNX13A124G, 
NASA Origins of Solar Systems Program grant NNX10AH40G via sub-award agreement 1312645088477, 
NASA Solar System Observations grant NNX16AD69G, 
BSF Grant Number 2012384,
as well as support from the Smithsonian 2015 CGPS/Pell Grant program.
DV and BTG also benefited by support from the European Union through ERC grant number 320964.




\label{lastpage}

\begin{thebibliography}{99}


\bibitem[Adams \& Bloch(2013)]{adablo2013} Adams, F.~C., \& Bloch, 
A.~M.\ 2013, ApJL, 777, L30 


\bibitem[Alcock et al.(1986)]{alcetal1986} Alcock, C., Fristrom, 
C.~C., \& Siegelman, R.\ 1986, ApJ, 302, 462 

\bibitem[Alonso et al.(2016)]{aloetal2016} Alonso, R., 
Rappaport, S., Deeg, H.~J., \& Palle, E.\ 2016, arXiv:1603.08823 

\bibitem[Antoniadou \& Veras(2016)]{antver2016} Antoniadou, K.~I., \& Veras, D.\ 2016, MNRAS, In Press, arXiv:1609.01734



\bibitem[Bear \& Soker(2013)]{beasok2013} Bear, E., 
\& Soker, N.\ 2013, New Astronomy, 19, 56 








\bibitem[Bonsor et al.(2011)]{bonetal2011} Bonsor, A., Mustill, 
A.~J., \& Wyatt, M.~C.\ 2011, MNRAS, 414, 930 

\bibitem[Bonsor \& Wyatt(2012)]{bonwya2012} Bonsor, A., \& 
Wyatt, M.~C.\ 2012, MNRAS, 420, 2990 

\bibitem[Bonsor et al.(2013)]{bonetal2013} Bonsor, A., Kennedy, 
G.~M., Crepp, J.~R., et al.\ 2013, MNRAS, 431, 3025 

\bibitem[Bonsor et al.(2014)]{bonetal2014} Bonsor, A., Kennedy, 
G.~M., Wyatt, M.~C., Johnson, J.~A., 
\& Sibthorpe, B.\ 2014, MNRAS, 437, 3288 




\bibitem[Chambers(1999)]{Chambers99} Chambers, J.~E.\ 1999, 
MNRAS, 304, 793 

\bibitem[Croll et al.(2016)]{croetal2016} Croll, B., Dalba, P.~A., Vanderburg, A., et al.\ 2016, Submitted to ApJL, arXiv:1510.06434 


\bibitem[Debes et al.(2011)]{debetal2011} Debes, J.~H., Hoard, D.~W., Wachter, S., Leisawitz, D.~T., \& Cohen, M.\ 2011, ApJS, 197, 38 

\bibitem[Debes et al.(2012)]{debetal2012} Debes, J.~H., Walsh, 
K.~J., \& Stark, C.\ 2012, ApJ, 747, 148 
%


\bibitem[Dufour et al.(2007)]{dufetal2007} Dufour, P., Bergeron, P., Liebert, J., et al.\ 2007, ApJ, 663, 1291

\bibitem[Dufour et al.(2012)]{dufetal2012} Dufour, P., Kilic, M., 
Fontaine, G., et al.\ 2012, ApJ, 749, 6 


\bibitem[Efroimsky \& Makarov (2013)]{EM2013} Efroimsky M., Makarov V.~V., 2013, ApJ, 764, 26 

\bibitem[Faedi et al.(2011)]{faeetal2011} Faedi, F., West, R.~G., Burleigh, M.~R., Goad, M.~R., \& Hebb, L.\ 2011, MNRAS, 410, 899 




\bibitem[Farihi et al.(2012)]{faretal2012} Farihi, J., 
G{\"a}nsicke, B.~T., Steele, P.~R., et al.\ 2012, MNRAS, 421, 1635 

\bibitem[Farihi et al.(2013)]{faretal2013} Farihi, J., G{\"a}nsicke, B.~T., \& Koester, D.\ 2013, Science, 342, 218 

\bibitem[Farihi(2016)]{farihi2016} Farihi, J. 2016, In Press New Astronomy Reviews, arXiv:1604.03092

\bibitem[Frewen \& Hansen(2014)]{frehan2014} Frewen, S.~F.~N., 
\& Hansen, B.~M.~S.\ 2014, MNRAS, 439, 2442 


\bibitem[Fulton et al.(2014)]{fuletal2014} Fulton, B.~J., Tonry, J.~L., Flewelling, H., et al.\ 2014, ApJ, 796, 114 

\bibitem[G{\"a}nsicke et al.(2006)]{gaeetal2006} G{\"a}nsicke, 
B.~T., Marsh, T.~R., Southworth, J., 
\& Rebassa-Mansergas, A.\ 2006, Science, 314, 1908 

\bibitem[G{\"a}nsicke et al.(2007)]{gaeetal2007} G{\"a}nsicke, 
B.~T., Marsh, T.~R., \& Southworth, J.\ 2007, MNRAS, 380, L35 

\bibitem[G{\"a}nsicke et al.(2008)]{gaeetal2008} G{\"a}nsicke, 
B.~T., Koester, D., Marsh, T.~R., Rebassa-Mansergas, A., 
\& Southworth, J.\ 2008, MNRAS, 391, L103 

\bibitem[G{\"a}nsicke(2011)]{gaensicke2011} G{\"a}nsicke, B.~T.\ 
2011, American Institute of Physics Conference Series, 1331, 211 

\bibitem[G{\"a}nsicke et al.(2012)]{gaeetal2012} G{\"a}nsicke, B.~T., Koester, D., Farihi, J., et al.\ 2012, MNRAS, 424, 333 

\bibitem[G\"{a}nsicke et al.(2016)]{gaeetal2016} G\"{a}nsicke, B.~T., Aungwerojwit, A., Marsh, T.~R. et al.\ 2016, In Press ApJL, arXiv:1512.09150

\bibitem[Gary et al.(2016)]{garetal2016} Gary, B.~L., Rappaport, S., Kaye, T.~G., Alonso, R., \& Hambsch, F.-J.\ 2016, Submitted to MNRAS, arXiv:1608.00026 

\bibitem[Gentile Fusillo et al.(2015)]{genetal2015} Gentile Fusillo, N.~P., G{\"a}nsicke, B.~T., \& Greiss, S.\ 2015, MNRAS, 448, 2260



\bibitem[Graham et al.(1990)]{graetal1990} Graham, J.~R., Matthews, 
K., Neugebauer, G., \& Soifer, B.~T.\ 1990, ApJ, 357, 216

\bibitem[Guo et al.(2015)]{guoetal2015} Guo, J., Tziamtzis, A., 
Wang, Z., et al.\ 2015, ApJL, 810, L17 

\bibitem[Gurri et al.(2016)]{guretal2016} Gurri, P., Veras, D., \& G{\"a}nsicke, B.~T.\ 2016, MNRAS, In Press, arXiv:1609.02563 

\bibitem[Henning \& Hurford (2014)]{HH2014} Henning W.~G., Hurford T., 2014, ApJ, 789, 30 



\bibitem[Hogan et al.(2009)]{hogetal2009} Hogan, E., Burleigh, M.~R., \& Clarke, F.~J.\ 2009, MNRAS, 396, 2074 



\bibitem[Jura(2003)]{jura2003} Jura, M.\ 2003, ApJL, 584, L91 


\bibitem[Jura et al.(2012)]{juretal2012} Jura, M., Xu, S., Klein, B., Koester, D., \& Zuckerman, B.\ 2012, ApJ, 750, 69 

\bibitem[Jura \& Young(2014)]{juryou2014} Jura, M., \& Young, E.~D.\ 2014, Annual Review of Earth and Planetary Sciences, 42, 45 



\bibitem[Kepler et al.(2015)]{kepetal2015} Kepler, S.~O., Pelisoli, I., Koester, D., et al.\ 2015, MNRAS, 446, 4078 

\bibitem[Kepler et al.(2016)]{kepetal2016} Kepler, S.~O., Pelisoli, I., Koester, D., et al.\ 2016, MNRAS, 455, 3413

\bibitem[Klein et al.(2010)]{kleetal2010} Klein, B., Jura, M., Koester, D., Zuckerman, B., \& Melis, C.\ 2010, ApJ, 709, 950 

\bibitem[Klein et al.(2011)]{kleetal2011} Klein, B., Jura, M., Koester, D., \& Zuckerman, B.\ 2011, ApJ, 741, 64 

\bibitem[Kleinman et al.(2013)]{kleetal2013} Kleinman, S.~J., Kepler, S.~O., Koester, D., et al.\ 2013, ApJS, 204, 5





\bibitem[Koester et al.(2014)]{koeetal2014} Koester, D., G{\"a}nsicke, B.~T., \& Farihi, J.\ 2014, A\&A, 566, A34 


\bibitem[Kunitomo et al.(2011)]{kunetal2011} Kunitomo, M., Ikoma, 
M., Sato, B., Katsuta, Y., \& Ida, S.\ 2011, ApJ, 737, 66 



\bibitem[Li et al.(2014)]{lietal2014} Li, G., Naoz, S., Kocsis, 
B., \& Loeb, A.\ 2014, ApJ, 785, 116 


\bibitem[Manser et al.(2016)]{manetal2016} Manser, C.~J., 
G{\"a}nsicke, B.~T., Marsh, T.~R., et al.\ 2016, MNRAS, 455, 4467 

\bibitem[Matsumura et al.(2013)]{matetal2013} Matsumura, S., Ida, 
S., \& Nagasawa, M.\ 2013, ApJ, 767, 129 

\bibitem[Melis et al.(2012)]{meletal2012} Melis, C., Dufour, P., 
Farihi, J., et al.\ 2012, ApJL, 751, L4 

\bibitem[Metzger et al.(2012)]{metetal2012} Metzger, B.~D., 
Rafikov, R.~R., \& Bochkarev, K.~V.\ 2012, MNRAS, 423, 505 

\bibitem[Mullally et al.(2008)]{muletal2008} Mullally, F., Winget, D.~E., Degennaro, S., et al.\ 2008, ApJ, 676, 573-583 

\bibitem[Mustill \& Villaver(2012)]{musvil2012} Mustill, A.~J., 
\& Villaver, E.\ 2012, ApJ, 761, 121 

\bibitem[Mustill et al.(2014)]{musetal2014} Mustill, A.~J., Veras, D., \& Villaver, E.\ 2014, MNRAS, 437, 1404

\bibitem[Nordhaus \& Spiegel(2013)]{norspi2013} Nordhaus, J., 
\& Spiegel, D.~S.\ 2013, MNRAS, 432, 500



\bibitem[Payne et al.(2013)]{payetal2013} Payne, M.~J., Deck, K.~M., Holman, M.~J., \& Perets, H.~B.\ 2013, ApJL, 775, L44

\bibitem[Payne et al.(2016)]{payetal2016} Payne, M.~J., Veras, D., Holman, M.~J., G\"{a}nsicke, B.~T.\ 2016, MNRAS, 457, 217



\bibitem[Raddi et al.(2015)]{radetal2015} Raddi, R., G{\"a}nsicke, 
B.~T., Koester, D., et al.\ 2015, MNRAS, 450, 2083 

\bibitem[Rafikov(2011a)]{rafikov2011a} Rafikov, R.~R.\ 2011a, ApJL, 
732, L3 

\bibitem[Rafikov(2011b)]{rafikov2011b} Rafikov, R.~R.\ 2011b, MNRAS, 
416, L55 

\bibitem[Rafikov \& Garmilla(2012)]{rafgar2012} Rafikov, R.~R., 
\& Garmilla, J.~A.\ 2012, ApJ, 760, 123 

\bibitem[Rappaport et al.(2016)]{rapetal2016} Rappaport, S., Gary, 
B.~L., Kaye, T., et al.\ 2016, Submitted to MNRAS, arXiv:1602.00740 

\bibitem[Raymond et al.(2010)]{rayetal2010} Raymond, S.~N., 
Armitage, P.~J., \& Gorelick, N.\ 2010, ApJ, 711, 772 

%

\bibitem[Redfield et al.(2016)]{redetal2016} Redfield, S., Farihi, J., Cauley, P.~W., et al.\ 2016, Submitted to ApJ, arXiv:1608.00549 




\bibitem[Sandhaus et al.(2016)]{sanetal2016} Sandhaus, P.~H., Debes, J.~H., Ely, J., Hines, D.~C., \& Bourque, M.\ 2016, ApJ In Press, arXiv:1604.03026 

\bibitem[Schatzman(1945)]{schatzman1945} Schatzman, E.\ 1945, Annales 
d'Astrophysique, 8, 143 



\bibitem[Staff et al.(2016)]{staetal2016} Staff, J.~E., De Marco, 
O., Wood, P., Galaviz, P., \& Passy, J.-C.\ 2016, MNRAS, In Press, arXiv:1602.03130 

\bibitem[Steele et al.(2011)]{steetal2011} Steele, P.~R., Burleigh, M.~R., Dobbie, P.~D., et al.\ 2011, MNRAS, 416, 2768 

\bibitem[Stone et al.(2015)]{stoetal2015} Stone, N., Metzger, 
B.~D., \& Loeb, A.\ 2015, MNRAS, 448, 188


\bibitem[Vanderburg et al.(2015)]{vanetal2015} Vanderburg, A., Johnson, J.~A., Rappaport, S., et al.\ 2015, Nature, 526, 546


\bibitem[Veras \& Armitage(2005)]{verarm2005} Veras, D., \& Armitage, P.~J.\ 2005, ApJL, 620, L111 




\bibitem[Veras \& Wyatt(2012)]{verwya2012} Veras, D., 
\& Wyatt, M.~C.\ 2012, MNRAS, 421, 2969 


\bibitem[Veras et al.(2013a)]{veretal2013a} Veras, D., Mustill, 
A.~J., Bonsor, A., \& Wyatt, M.~C.\ 2013a, MNRAS, 431, 1686 


\bibitem[Veras et al.(2014a)]{veretal2014a} Veras, D., Leinhardt, 
Z.~M., Bonsor, A., G{\"a}nsicke, B.~T.\ 2014a, MNRAS, 445, 2244 

\bibitem[Veras et al.(2014b)]{veretal2014b} Veras, D., Jacobson, 
S.~A., G{\"a}nsicke, B.~T.\ 2014b, MNRAS, 445, 2794 

\bibitem[Veras et al.(2014c)]{veretal2014c} Veras, D., Shannon, A., 
G{\"a}nsicke, B.~T.\ 2014c, MNRAS, 445, 4175

\bibitem[Veras et al.(2014d)]{veretal2014d} Veras, D., Evans, N.~W., 
Wyatt, M.~C., \& Tout, C.~A.\ 2014d, MNRAS, 437, 1127 

\bibitem[Veras \& G\"{a}nsicke(2015)]{vergae2015} Veras, D., G\"{a}nsicke, B.~T.\ 2015, MNRAS, 447, 1049 

\bibitem[Veras et al.(2015a)]{veretal2015a} Veras, D., Leinhardt, 
Z.~M., Eggl, S., \& G\"{a}nsicke, B.~T.\ 2015a, MNRAS, 451, 3453. 

\bibitem[Veras et al.(2015b)]{veretal2015b} Veras, D., Eggl, S., 
\& G\"{a}nsicke, B.~T.\ 2015b, MNRAS, 451, 2814.  

\bibitem[Veras et al.(2015c)]{veretal2015c} Veras, D., Eggl, S., G\"{a}nsicke, B.~T.\ 2015c, MNRAS, 452, 1945 

\bibitem[Veras(2016a)]{veras2016} Veras, D.\ 2016a, RSOS, 3:150571

\bibitem[Veras(2016b)]{veras2016b} Veras, D.\ 2016b, MNRAS In Press, arXiv:1608.07580 

\bibitem[Veras et al.(2016a)]{veretal2016a} Veras, D., Mustill, A.~J., G\"{a}nsicke, B.~T. et al.\ 2016a, In Press, MNRAS, arXiv:1603.00025

\bibitem[Veras et al.(2016b)]{veretal2016b} Veras, D., Marsh, T.~R., G\"{a}nsicke, B.~T. \ 2016b, Submitted to MNRAS

\bibitem[Veras et al.(2016c)]{veretal2016c} Veras, D., Carter, P.~J., Leinhardt, Z.~L., G\"{a}nsicke, B.~T. \ 2016c, Submitted to MNRAS

\bibitem[Villaver \& Livio(2009)]{villiv2009} Villaver, E., 
\& Livio, M.\ 2009, ApJL, 705, L8

\bibitem[Villaver et al.(2014)]{viletal2014} Villaver, E., Livio, 
M., Mustill, A.~J., \& Siess, L.\ 2014, ApJ, 794, 3



\bibitem[Xu et al.(2013)]{xuetal2013} Xu, S., Jura, M., Klein, B., Koester, D., \& Zuckerman, B.\ 2013, ApJ, 766, 132 

\bibitem[Xu et al.(2014)]{xuetal2014} Xu, S., Jura, M., Koester, D., Klein, B., \& Zuckerman, B.\ 2014, ApJ, 783, 79 


\bibitem[Xu et al.(2016)]{xuetal2016} Xu, S., Jura, M., Dufour, P., \& Zuckerman, B.\ 2016, In Press ApJL, arXiv:1511.05973


\bibitem[Wilson et al.(2015)]{wiletal2015} Wilson, D.~J., 
G{\"a}nsicke, B.~T., Koester, D., et al.\ 2015, MNRAS, 451, 3237 




\bibitem[Zuckerman et al.(2003)]{zucetal2003} Zuckerman, B., 
Koester, D., Reid, I.~N., H\"{u}nsch, M.\ 2003, ApJ, 596, 477 

\bibitem[Zuckerman et al.(2010)]{zucetal2010} Zuckerman, B., Melis, 
C., Klein, B., Koester, D., \& Jura, M.\ 2010, ApJ, 722, 725 


\end{thebibliography}
\end{document}